\documentclass[a4paper]{IEEEtran}
\usepackage{epsf}
\newcommand{\ul}[1]{\underline{#1}}

\newcommand{\eps}{\epsilon}

\newcommand{\bea}{\begin{eqnarray}}

\newcommand{\eea}{\end{eqnarray}}

\newcommand{\rmc}{{\rm c}}

\newcommand{\rmd}{{d}}

\newcommand{\rmf}{{f}}

\newcommand{\rmj}{{\rm j}}

\newcommand{\rmr}{{\rm r}}

\newcommand{\rmw}{{\rm w}}

\newcommand{\Beta}{{\rm B}}

\begin{document}


\title{Measurement Method for Evaluating the Probability Distribution of the Quality Factor of Mode-Stirred Reverberation Chambers}

\author{
{\it	Luk R. Arnaut\footnote{now with George Green Institute of Electromagnetics Research, University of Nottingham, Nottingham NG7 2RD, U.K.}, Mihai I. Andries\footnote{now with Institut Francais des Sciences et Technologies des Transports (IFFSTAR), de l'Am\'{e}nagement et des R\'{e}seaux, Villeneveuve d'Ascq, France}, J\'{e}r\^{o}me Sol, Philippe Besnier} 
\\
Institute of Telecommunications Research, National Institute of Applied Sciences (INSA),
Rennes, France\\
(\{luk.arnaut,mihai.andries,jerome.sol,philippe.besnier\}@insa-rennes.fr)
}
\date{\today}

\maketitle

\begin{abstract}
An original experimental method for determining the empirical probability distribution function (PDF) of the quality factor ($Q$) of a mode-stirred reverberation chamber is presented.  
Spectral averaging of S-parameters across a relatively narrow frequency interval at a single pair of locations for the transmitting and receiving antennas is applied to estimate the stored and dissipated energy in the cavity, avoiding the need for spatial scanning to obtain spatial volume or surface averages. 
The effective number of simultaneously excited cavity modes per stir state, $M$, can be estimated by fitting the empirical distribution to the parametrized theoretical distribution.
The measured results support a previously developed theoretical model for the PDF of $Q$ and show that spectral averaging over a bandwidth as small as a few hundred kHz is sufficient to obtain accurate results.
\end{abstract}
Keywords: Maximum likelihood estimation, method of moments, mode-stirred reverberation chamber, probability distribution, quality factor.

\section{Introduction \label{sec:intro}}
The propagation of uncertainties or probability density functions (PDFs) is a key aspect in the probabilistic characterization of complex electromagnetic (EM) systems and environments. This requires the standard deviation or full PDF for each fluctuating input parameter to be specified, respectively. Techniques such as polynomial chaos expansion or stochastic collocation then enable the corresponding uncertainty of an output quantity of interest to be calculated \cite{chau1,bonn1}.

Mode-tuned and mode-stirred reverberation chambers (MT/MSRCs) offer an experimental facility for emulating dynamic multipath propagation. Several of its intrinsic EM wave parameters for the interior random fields require statistical characterization, e.g. impedance, wave number, etc. \cite[Sec. 3.5]{arnaAEU}, \cite{serr1}. 
Since a MT/MSRC is a resonant environment, its quality ($Q$) factor is of fundamental importance and requires accurate statistical characterization. Since $Q$ is a field-related quantity, it exhibits quasi-random fluctuations during the stir process: 
\bea
Q(\omega, \tau) \stackrel{\Delta}{=} \omega \frac{U(\omega, \tau) }{ P_\rmd(\omega, \tau)}
\label{eq:defQ}
\eea
where $U$ is the EM stored energy and $P_\rmd$ is the dissipated power, both evaluated at frequency $f=\omega/(2\pi)$ and stir state $\tau$.
In a recent paper \cite{arnaTEMC2013}, we derived a PDF $f_Q(q)$ of $Q$ for an assumed ideal (i.e., statistically isotropic, homogeneous, unpolarized and fully incoherent) cavity field, where field randomization is generated by a perfectly efficient mode stirrer. This assumption allows for the modelling the PDF of $Q$ from the ratio of a $\chi^2_{6M}$ distributed $U$ and a $\chi^2_{4M}$ distributed $P_\rmd$, where $M$ is the number of degrees of freedom (NDoF) of the stirring process.

Although $U$ and $P_\rmd$ are in principle directly measurable (observable) quantities, the experimental determination of $U$ and $P_\rmd$ and their PDFs individually to obtain $f_Q(q)$ is not straightforward: $U$ and $P_\rmd$ are defined as integrals of nonlocal (i.e., spatially extended) quantities. For random field exhibiting complicated, i.e., quasi-random spatial distributions (maps) across volume and surface, this is in fact not achievable except for extremely dense sampling. At best, $U$ and $P_\rmd$ can only be estimated by aggregation of sampled values, across scanned point locations on a structured or unstructured grid and subjected to local averaging effects. Furthermore, accurate measurement of fields across extended regions and near surfaces meets several difficult in practice, viz., cabling effects, near-field coupling between antennas and conducting boundaries requiring probe correction, etc. Finally, mechanical scanning (repositioning) is considerably more cumbersome and time consuming than e.g. frequency scanning.

Statistics of cavity resonances and their fluctuating spectral widths has also been studied from a `microscopic' point of view \cite{gros1}. However, the measurement of individual resonances in practical MT/MSRCs beyond the first few ones is limited by the rapidly increasing modal overlap with increasing frequency and instrumentation limitations.

In this paper, it is shown that, as an alternative to spatial scanning, measurements across a relatively narrow frequency interval allow for a reasonably accurate and effective estimation of the total volumetric stored energy and surfacial dissipated power, at an arbitary stir state. 
These values can then be randomized by mode stirring to yield empirical probability distributions of $U$, $P_\rmd$ and, hence, the `macroscopic' $Q$ for comparison with the model in \cite{arnaTEMC2013}.

\section{Measurement Method\label{sec:method}}
Since $U$ is a volumetric quantity within the cavity volume that depends on the vectorial field,
we aim to estimate the volumetric stored energy $U$ based on measurements of Cartesian field components along three mutually orthogonal directions, {\em at a single spatial location} and at relatively large distances from the cavity boundaries (at least half a wavelength at the lowest measurement frequency) in order to avoid the aforementioned practical difficulties.
The spatial density of the stored electric and magnetic energy densities $U_e$ and $U_m$ can be estimated from measurements of the 3-D EM field as
\bea
U_e (\ul{r}) &=& \frac{\eps_0}{2} (|E_x(\ul{r}) |^2 + |E_y(\ul{r}) |^2 + |E_z(\ul{r}) |^2)\\
U_m (\ul{r}) &=& \frac{\mu_0}{2} (|H_x(\ul{r}) |^2 + |H_y(\ul{r}) |^2 + |H_z(\ul{r}) |^2).
\eea
At sufficiently high frequencies, $|\ul{E}|$ and $|\ul{H}|$ in the radiation zone 
satisfy $U_e=U_m$ in this region. In the vicinity of a rectilinear antenna, $U_e > U_m$ \cite{vanb2}. More generally, in the near field of the transmitting antenna, the local received power contains a reactive (imaginary) component that is governed by he imbalance between $U_e(\ul{r})$ and $U_m(\ul{r})$, i.e., $P = P_\rmf + P_\rmd + \rmj \omega (\overline{U}_e - \overline{U}_m)$, where the overbar denotes averaging.
For simultaneous excitation of multiple modes in overmoded regime, incoherent superposition can be applied, whereby the electric and magnetic energies are added rather than their originating complex fields. The local density of the total energy is then $U(\ul{r})=U_e(\ul{r})+U_m(\ul{r})$.
Across the cavity volume\footnote{In the near zone of the transmitting antenna, the deterministic line-of-sight component of the field has an amplitude that decreases as (a sum of components with) $(kr)^{-n}$, $n>1$ In the short wavelength limit, we neglect the resulting change to the total interior energy, i.e., we assume that all energy at each location inside the cavity is perfectly stirred.} $V$, 
the total energy is then
\bea
U = \int_V U(\ul{r}) {\rmd}V = \int_V \left [ U_e (\ul{r}) + U_m (\ul{r}) \right ] {\rm d}V
.
\eea

On the other hand, determining the dissipated energy would requires a precise measurement of boundary field, near flat as well as angled or irregular surfaces, where its statistical inhomogeneity is larger. 
While the EM boundary conditions near a PEC boundary dictate that the normal electric field is the dominant field component, for calculating the power dissipated by the walls, pairs of tangential components of $\ul{E}$ and $\ul{H}$ are required in order to evaluate the energy flow into the boundary.
It could be argued that the surfacial dissipated power $P_\rmd$ is governed by the interior modal structure or by waves impinging on the lossy boundary from within the cavity, hence its distribution is also governed by interior orthogonal field components, averaged over pairs of such components to be representative of irregular cavity walls in overmoded conditions.
When using CW excitation, however, a pragmatic approach is to note that the total power loss (i.e., dissipated in the cavity walls as well as in the antennas and cables, radiating through apertures, etc.) equals the transmitted power, i.e., $P_\rmd = P_0(1-|S_{11}|^2)$. Here, $P_0$ is the nominal output power.
Naturally, this measurement of the total dissipated power yields a good approximation of $P_\rmd$ for cavity wall losses only if the latter is the dominant loss mechanism, i.e., at sufficiently high frequencies, where antenna losses, cabling, (small) aperture effects, etc., are negligibly small.

In the overmoded regime, since the random field is statistically homogeneous, isotropic, uniformly
unpolarized, and reciprocal, the stored and dissipated energies can be deduced from sets of triplets of S-parameters ($S_{21}$, $S_{11}$, $S_{22}$) measured using a vector network analyzer. To this end, for a fixed $\tau$ and with $x,y,z$ forming an arbitrarily oriented triplet of mutually orthogonal\footnote{In overmoded regime, this requirement can be relaxed to due statistical isotropy and homogeneity, in which case three measurements at three arbitrary positions with arbitrary orientations are sufficient.} 
directions at $\ul{r}$, the value of $U$ can be estimated as 
\bea
U &=& \int_V u(\ul{r}) dV 
\equiv V \langle U(\ul{r}) \rangle_V 
= \frac{\eps_0 V}{2} \left \langle \sum_{\alpha=x,y,z} |E_\alpha|^2 \right \rangle_V\nonumber\\
&\propto&
\sum_{\alpha=x,y,z} \left \langle \frac{ |S_{21,\alpha}|^2}{1-|S_{22,\alpha}|^2} \right \rangle_V 
.
\label{eq:Utot}
\label{eq:U_final}
\eea  
To determine the volume average $\langle \cdot \rangle_V$, measurements at multiple locations are strictly required. This would involve a time-consuming and limited process of spatial scanning. 
This can be circumvented by replacing the volume average approximately with a spectral average.
This approximation -- which is in essence an application\footnote{Strictly, this requires a test for first-order ergodicity of the field intensity to hold, such that
volume averages can be replaced by spectral averages across a relatively narrow frequency band, or vice versa in case of frequency stirring.
Since we aim to generate a set of random values for $U$ and $P_\rmd$, we must maintain the volumetric/surfacial estimates to remain random variables after estimation, in order to enable statistical analysis. Although narrowband excitation is needed in order for the concept and definition of quality factor to remain meaningful, it appears that a PDF for a spectrally averaged $Q$ follows automatically from our method as well.} 
of the ergodic theorem -- holds provided that the averaging bandwidth is not too small. In this case, a sufficiently large number of independent sample values of the S-parameters is generated that enables an accurate estimation of the averages $\langle |S_{ij}|^2 \rangle_V$. 
Hence, in overmoded regime, the net energy received by the Rx antenna or probe, expressed with reference to the input power at the calibration (reference) plane at the connector of the Tx antenna, $P_T(\tau)$, is
\bea
U \propto \frac{P_0 V}{\eta_0} \sum_{\alpha=x,y,z} \left \langle \frac{ |S_{21,\alpha}|^2}{1-|S_{22,\alpha}|^2} \right \rangle_f 
.
\eea
This expression uses the free-space impedance $\eta_0$ as the average wave impedance, but neglects the statistical variations of $\eta(\tau)$ \cite{arnaAEU}, \cite{serr1}.
Similarly, the power dissipated in the cavity walls with interior surface area $S$ is estimated as the average net transmitted (forward injected) power
\bea
P_\rmd &=& \int_S P_\rmd(\ul{r}_S) dS = P_T \\
&\propto& 
\sum_{\alpha=x,y,z} \left \langle 1-|S_{11,\alpha}|^2 \right \rangle_f 
.
\label{eq:Pd_final}
\eea
Combining (\ref{eq:U_final}) and (\ref{eq:Pd_final}) finally yields
\bea
Q \propto \frac{\omega V}{S} \frac{\sum_{\alpha=x,y,z} \left \langle { |S_{21,\alpha}|^2} / \left ( {  1-|S_{22,\alpha}|^2 } \right ) \right \rangle_f }
{\sum_{\alpha=x,y,z} \left \langle 1-|S_{11,\alpha}|^2 \right \rangle_f }
.
\label{eq:Q_final_freqavg}
\eea

These expressions apply at a single ($n$th) arbitrary stir state $\tau$, acquiring $N$ values $q_s\equiv Q(\tau)$ during the stir process. These randomly fluctuating values determine a sample PDF and sample statistics of the random $Q$. If the set contains a sufficiently large number of independent stir states, then the sample PDF of $Q$ may be comparable to the theoretical ensemble PDF. 

\section{Theoretical PDF of $Q$}
For the comparison with experimentally determined $f_Q(q)$, we briefly summarize the main relevant results from \cite{arnaTEMC2013}. 
For an idealized random cavity field,
the PDF of $Q$ is found to be {\it a priori} calculable as the Fisher--Snedecor $F(2r,2s)$ distribution
\bea
f_Q(q) =
\frac{1}{\Beta(r,s)} \left ( \frac{h \thinspace V}{\mu_{\rmw,\rmr} \delta_\rmw S}\right )^s \frac{q^{r-1}}{\left ( q + \frac{h \thinspace V}{\mu_{\rmw,\rmr} \delta_\rmw S} \right )^{r+s}}
.
\label{eq:PDFQ_final_freqavg}
\eea
Here, $\mu_{\rmw,\rmr}$ and $\delta_{\rm w}$ are the relative permeability and skin depth of the cavity walls, respectively, B$(r,s)$ is a complete beta function associated with a $\chi^2_{2r}$ PDF for the energy stored in $V$ and a $\chi^2_{2r}$ PDF for the power dissipated across $S$, whereas $h$ is an ensemble averaged shape factor of the stirred cavity, weighted over the contributing eigenmodes $\ul{\phi}$ and defined by
$
h
\stackrel{\Delta}{=} {2~\langle \langle |\ul{\phi}(\ul{r})|^2 \rangle_V \rangle}/{\langle \langle |\ul{\phi}(\ul{r}_S)|^2 \rangle_S \rangle}
$,
with $h=1$ for a rectangular cavity \cite{arnaTEMC2013} and for an irregularly shaped cavity \cite{denn1}, where $\langle \cdot \rangle$ denotes ensemble (i.e., stir) averaging.
For comparison with experimental data, the following equivalent form is useful \cite{arnaTEMC2013}: 
\bea
f_Q(q) =
\frac{1}{\Beta(r,s)} \left ( \frac{s-1}{r} \langle Q \rangle \right )^s \frac{q^{r-1}}{\left ( q + \frac{s-1}{r} \langle Q \rangle \right )^{r+s}}
.
\label{eq:PDFQ_selfsuff_mu}
\eea
Here, $s=2M$ and $r=3M$ are parameters and the NDoF $M$ may be interpreted in the asymptotic limit (large $M$) as the number of simultaneously excited cavity modes per stir state.
In this way, $M$ is a single parameter of the distribution for mean-normalized values of $Q$.
On the other hand, formally allowing for $s$ and $r$ to be two {\em independent\/} parameters of the PDF (i.e., not necessarily related as $r/s=3/2$) offers more flexibility in matching empirical to theoretical distributions, for example as a result of a disproportionally large number of DoFs for the energy dissipation ($s > 2M$) compared to energy storage ($s < 3M$) in nonideal scenarios.

\subsection{Estimation of $M$}
\subsubsection{Prior Estimation}
If no experimental data is available then, in a first approximation, $M$ can be estimated as the average number of excited modes per stir state, based on the density of modes multiplied by the average coherence bandwidth $\langle \delta f \rangle = \langle f / Q \rangle \simeq f / \langle Q \rangle$ to yield
\bea
M \simeq \frac{dN}{df} \langle \delta f \rangle
.
\eea
For CW excitation, its value can be estimated\footnote{neglecting a LF correction due to the edges of the cavity} based on the asymptotic Weyl law for the average density of cavity modes, as \cite[eqs. (63)-(64)]{arnaTEMC2013}
\bea
M(f) = \frac{b + \sqrt{b^2-3b}}{3}
\label{eq:estM}
\eea
with
\bea
b \stackrel{\Delta}{=} \frac{3}{2} M_\infty (f) = {8\pi} \frac{\mu_{\rmw,\rmr} \delta_\rmw S}{h} \left ( \frac{f}{\rmc} \right )^3
= \frac{12\pi V} {Q_\infty} \left ( \frac{f}{\rmc} \right )^3,
\eea
where $M_\infty = M(f\rightarrow +\infty)$ is the asymptotic average number of excited modes and
\bea
Q_\infty \stackrel{\Delta}{=} \frac{3 \thinspace h \thinspace V}{2 \mu_{\rmw,\rmr} \delta_\rmw S}
 \label{eq:avgQ_asymp}
\eea
is the asymptotic average $Q$ factor (``composite $Q$'').
The value determined by (\ref{eq:estM}) defines the prior estimate $M_{prior}$. 
Note that $M_{prior}$ is itself a spectral average (effective) number, because detailed spectral fluctuations and associated variations in modal weighting are neglected in Weyl's law and in its low-frequency first-order generalization.

\subsubsection{Posterior Estimation}
\paragraph{Method of Moments}
Based on the measured S-parameters and derived values $Q(\tau)$, a point estimate of $M$ can be obtained using the statistical {method of moments} (MoM).
To this end, sample values of the coefficient of variation of $\nu_Q \stackrel{\Delta}{=} \sigma_Q / \mu_Q$ can be used, whose asymptotic population value is
$
\nu_Q \rightarrow \sqrt{{5}/{6M}}
$ for $M\rightarrow +\infty$ \cite[Eq. (20)]{arnaTEMC2013},
with which
\bea
M \simeq {\frac{5}{6 n^2_Q}} \label{eq:M_MoM_empir}
\eea
where $n_Q \stackrel{\Delta}{=} s_Q/m_Q$ is the sample estimate of $\nu_Q$. At lower frequencies,
more accurate estimates than (\ref{eq:M_MoM_empir}) are obtained by inverting the general expression for $\nu_Q$, which is also valid for  $M \not \gg 1$ \cite[Eq. (17)]{arnaTEMC2013}.

\paragraph{Maximum Likelihood Estimation}
Alternatively, one may estimate $M$ from the theoretical cumulative distribution (CDF) \cite[eqs. (48)-(50)]{arnaTEMC2013}
\bea
F_Q(q) = 1-I_\xi(2M,3M)
\label{eq:CDFQ_theo}
\eea
where $I_{\xi}(2M,3M)$ is the regularized incomplete beta function with
\bea
\xi \stackrel{\Delta}{=} \left ( 1 + \frac{3M}{2M-1} \frac{q}{\langle Q \rangle} \right )^{-1}
.
\label{eq:CDFQ_theo_xi}
\eea
In this method, the experimentally determined ranked values of $Q(\tau,f)$ are used to fit the empirical CDF to the theoretical CDF, by optimizing the value of $M$ that yields a best fit. This defines the {maximum likelihood estimation} (MLE) of $M$, which has been used for other applications of parameter extraction in EM, cf., e.g., \cite{arnaICA}. 
Specifically, the fit between both CDFs is optimized by minimizing their mean squared difference through adjustment of $M$ and $a=C/\langle Q \rangle$ as free parameters of the modelled CDF, where $C$ is the normalization constant of the CDF.
This was done here using the Nelder--Mead method, which performs an unconstrained nonlinear minimization of the sum of squared residuals. 
To monitor the quality of the fit (and, hence, the accuracy of the thus estimated values of $M$),
the least-square error (LSE) correlation coefficient $\varrho$
and its associated coefficient of determination $R=\varrho^2$ provide measures of the quality of the fit of the CDFs.
The deviation of the normalization coefficient for the empirical CDF from its ideal unit value, $a$, provides a alternative measure of fit between the CDFs.

\subsection{Spectral Averaging vs. Spatial Averaging}
Although the frequency averages $\langle U(\tau) \rangle_f$ and $\langle P_\rmd(\tau) \rangle_f$ estimate $\langle U(\tau) \rangle_V$ and $\langle P_\rmd(\tau) \rangle_V$, the calculated $Q$ only applies across the frequency band of overlapping (coupled) modes excited through a {\em single\/} CW frequency, by definition of $Q$  in (\ref{eq:defQ}). 
In overmoded regime, this band of coupled modes is wider than the individual modal bandwidth, because mode density increases faster $(\propto f^2)$ than the decrease of the modal spectral width ($\propto f^{-1/2}$), leading to progressively larger average modal overlap with increasing frequency.
Nevertheless, this coupling bandwidth is still much smaller than the typical interval for frequency averaging used to calculate $\langle U(\tau) \rangle_f$ and $\langle P_\rmd(\tau)\rangle_f$, the latter being typical tens of modal bandwidths.

\subsection{Uncertainty of $M$}
Similar to the issue of the equivalent number of statistically independent (or even merely uncorrelated) stir states, the number of participating modes $M$ per stir state for CW excitation is not a sharply defined quantity. Indeed, cavity resonance frequencies that do not exactly match the CW excitation frequency will only be partially excited (weighted), with a weight coefficient that is proportional to the amplitude of the Lorentz spectrum at the specific $f$.
Thus, $M$ is at best an approximate effective number that has, in general, a fractional value. In any case, $M$ is useful as a (fluctuating) parameter of the distribution.

\section{Experimental Results}
Measurements were conducted in the reverberation chamber at INSA Rennes, France, measuring $L\times W \times H =8.7\times 3.7\times 2.9$ m${^3}$. The chamber comprises a paddle wheel mechanical mode stirrer consisting of six aluminium blades of dimensions $1 \times 0.75 \times 0.005$ m$^3$.
The stirrer was rotated across one full turn in $N_s=1400$ equiangular steps. At each stir state, after allowing for a dwell time of $8$ s to reach mechanical steady state, $40000$ triplets of complex S-parameters were recorded (one set per frequency) using two nominally identical broadband dipole antennas operating between $0.1$ and $6.1$ GHz in steps\footnote{Two frequency sweeps with spacing of $300$ kHz each were interleaved in post-processing, to surpass the capacity of the internal memory the VNA.} of $150$ kHz, after which the stirrer was moved to its next angular state and the measurement repeated.

\begin{figure}[!ht] \begin{center} \begin{tabular}{c}
\ \epsfxsize=8.6cm \hspace{-0.6cm}
\epsfbox{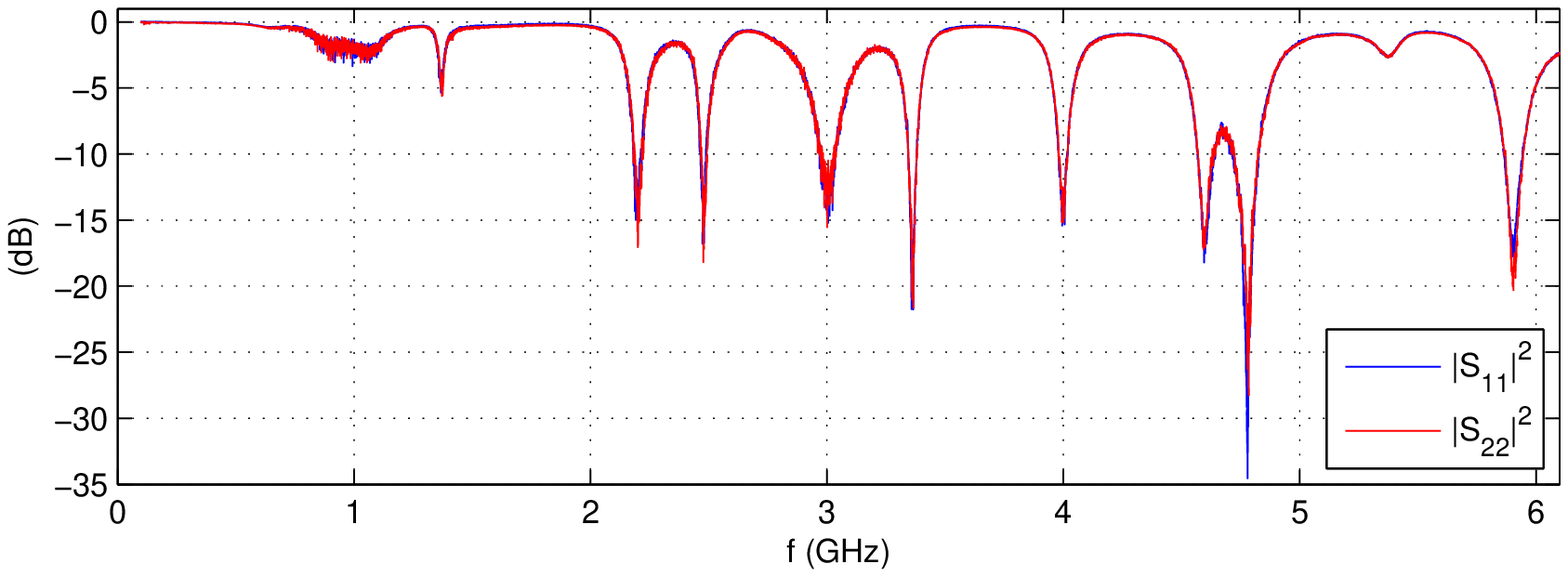}\ \\
(a)\\
\ \epsfxsize=8.6cm \hspace{-0.6cm}
\epsfbox{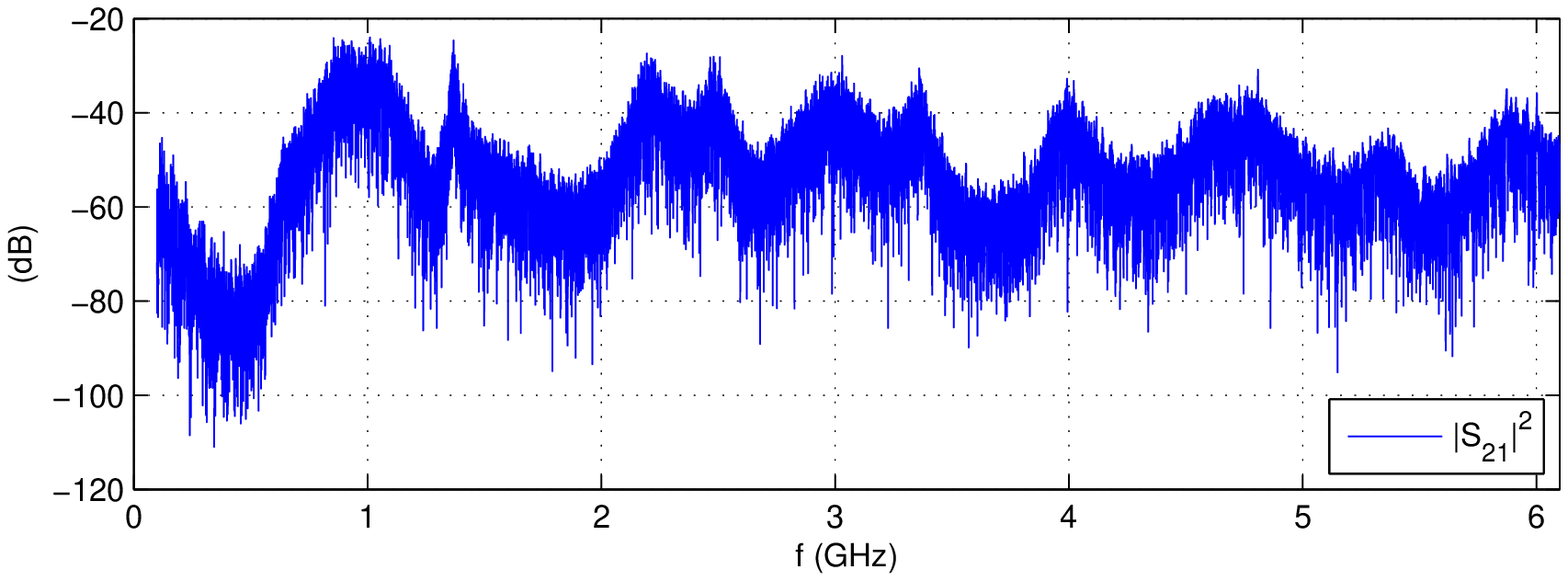}\ \\
(b)\\
\ \epsfxsize=8.6cm \hspace{-0.6cm}
\epsfbox{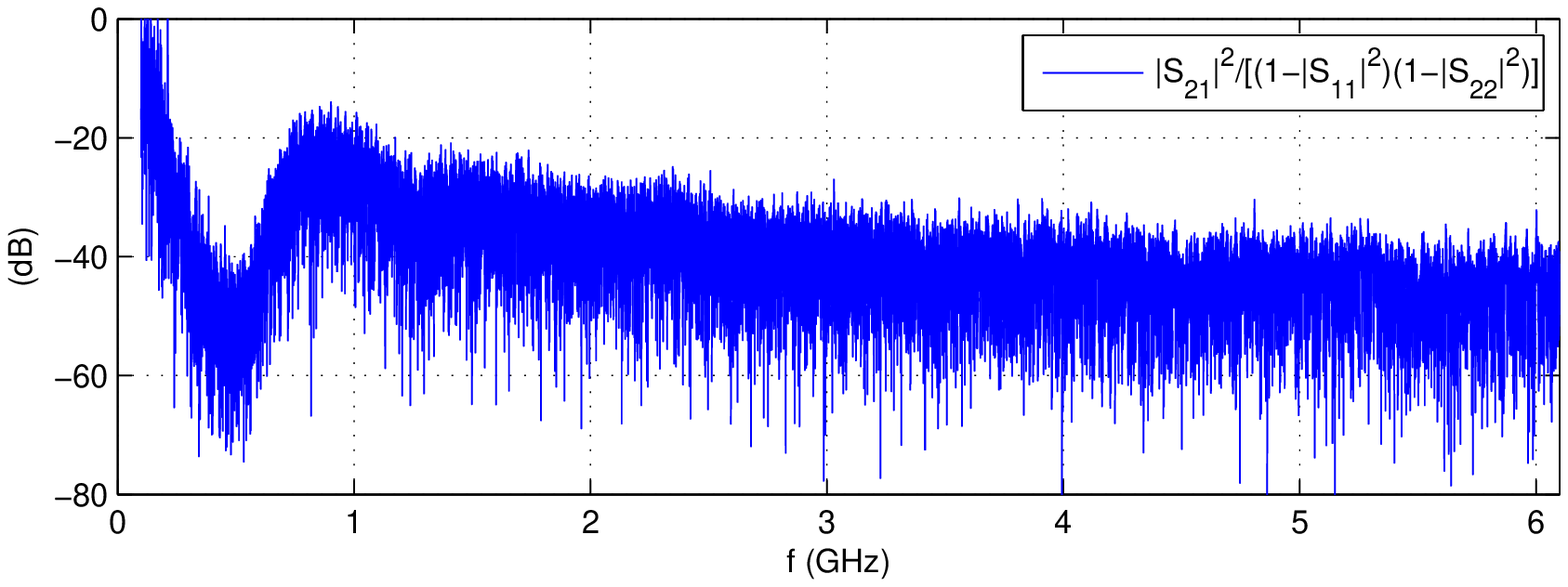}\ \\
(c)\\
\end{tabular}
\end{center}
{
\caption{\label{fig:Sparamsdipant} \small
Typical frequency responses of S-parameters of nominally identical broadband dipole antennas at ports 1 and 2, measured inside the MTRC at one arbitrary stir state.}
}
\end{figure}

 It is verified that the reflections $S_{11}(f)$ for both antennas are nearly identical (Fig. \ref{fig:Sparamsdipant}(a)). Hence, in principle only one pair of S-parameters ($S_{21}(f),S_{11}(f)$) needs to be measured and recorded, thus reducing measurement time and capacity requirements.
The strong impedance mismatch (weak coupling) of the antennas across several subbands was not found to affect the distribution of $Q$, as will be shown.
The net transmitted power between the antennas (Fig. \ref{fig:Sparamsdipant}(c)) exhibits a more monotonic average decrease with frequency compared to the raw $|S_{21}(f)|^2$ (Fig. \ref{fig:Sparamsdipant}(b)).

By spectral averaging of the $|S_{ij}|^2$ across a band of $n$ frequencies per stir state, $1400$ sample values $q_s$ were obtained in accordance with (\ref{eq:Q_final_freqavg}).
Thus, $40000/n$ empirical CDFs could be generated across the spectrum, each CDF formed by $1400$ sample values of $Q$ representative of this frequency band. 
The sample mean and standard deviation of the $1400$ values were used to derive $40000/n$ associated MoM estimates of $M$ based on (\ref{eq:M_MoM_empir}) and corresponding MLE estimates MLE by matching each empirical CDF to the theoretical CDF (\ref{eq:CDFQ_theo})--(\ref{eq:CDFQ_theo_xi}) through curve fitting using the Nelder--Mead algorithm, yielding $40000/n$  alternative estimates of $M$ and $C/\langle Q \rangle$.

\subsection{Estimation of $M$}
\begin{figure}[!ht] \begin{center} \begin{tabular}{c}
\ \epsfxsize=8.6cm \hspace{-0.6cm}
\epsfbox{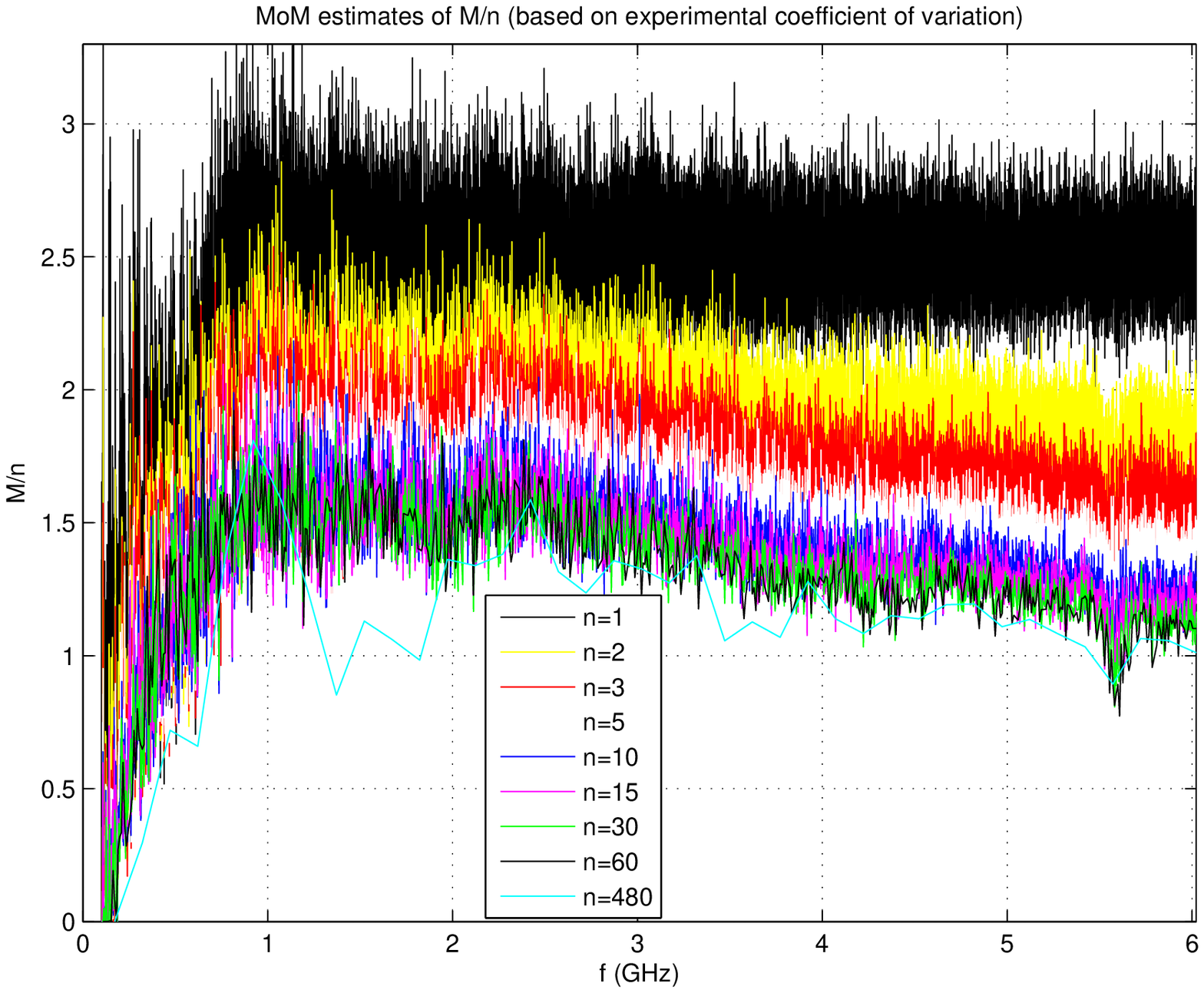}\ \\
(a)\\
\ \epsfxsize=8.6cm \hspace{-0.6cm}
\epsfbox{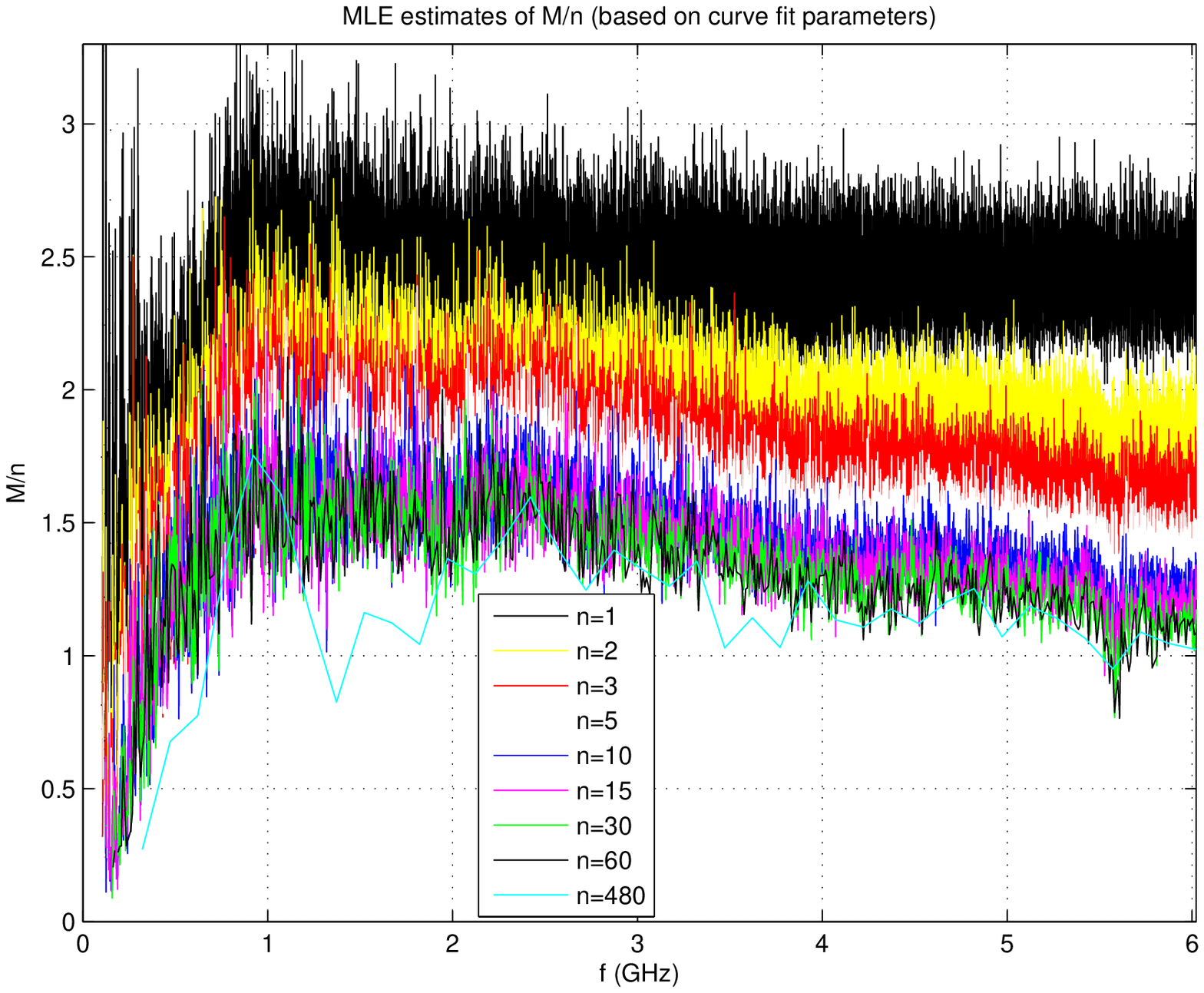}\ \\
(b)\\
\end{tabular}
\end{center}
{
\caption{\label{fig:Mestimates_MoM_MLE} \small
Estimated values of degrees of freedom parameter $M$ as a function of frequency, based on $n$-point frequency averaging at each one of $1400$ data points $q$ for $Q$, using (a) MoM estimation and (b) MLE point estimation.}
}
\end{figure}

\begin{figure}[!ht] \begin{center} \begin{tabular}{c}
\ \epsfxsize=8.6cm \hspace{-0.6cm} 
\epsfbox{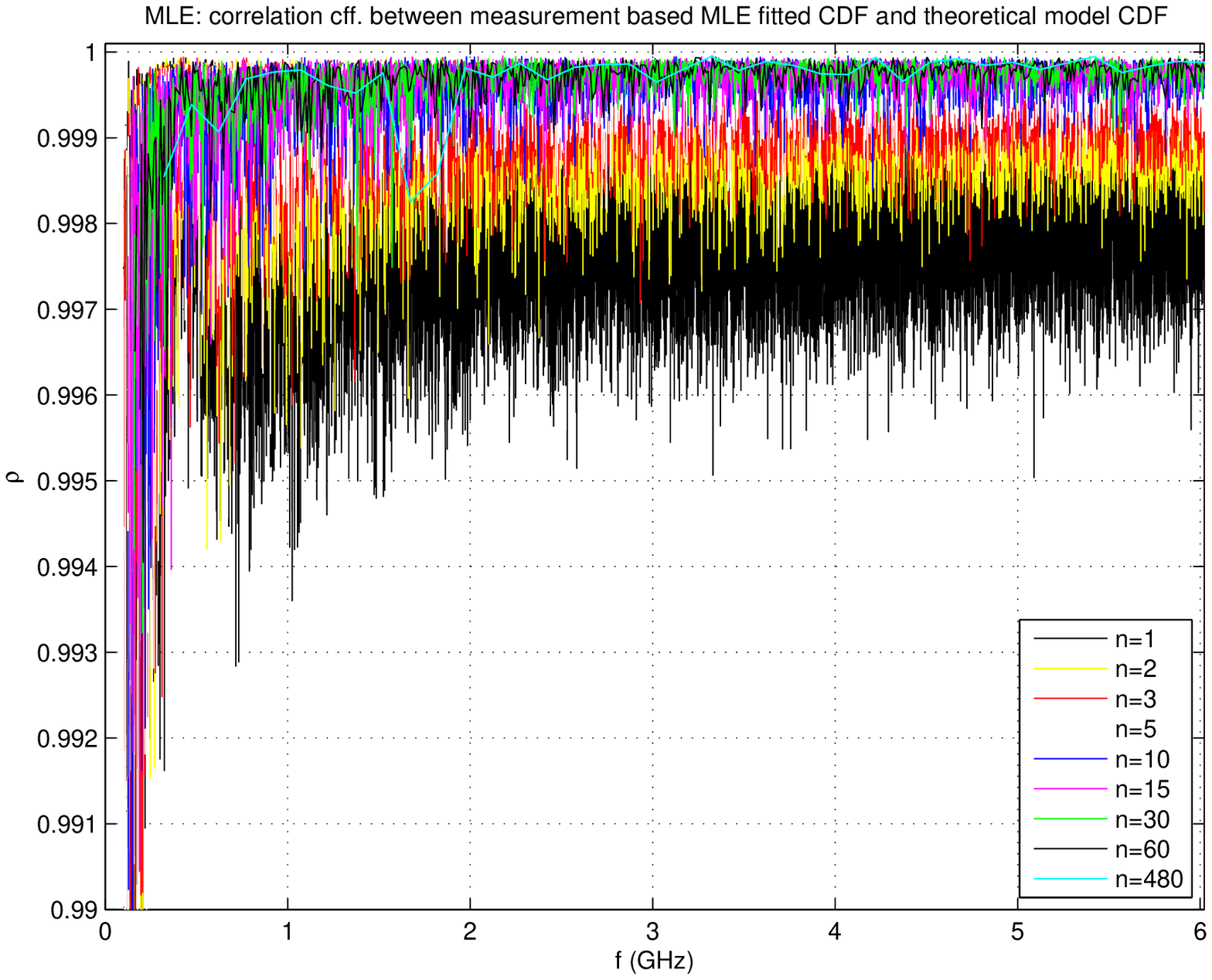}\ \\
(a)\\
\ \epsfxsize=8.6cm \hspace{-0.6cm} 
\epsfbox{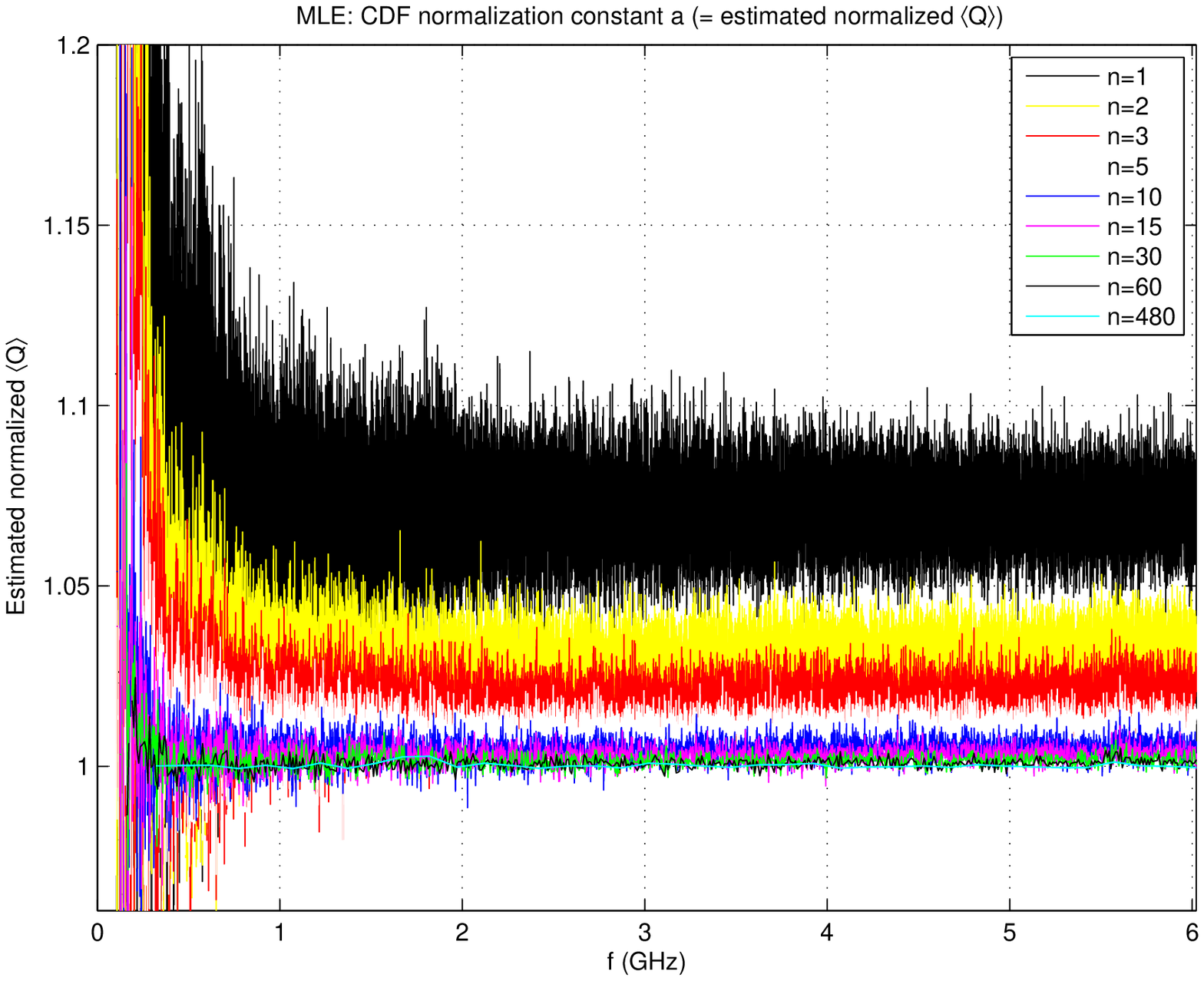}\ \\
(b)\\
\end{tabular}
\end{center}
{
\caption{\label{fig:Mestimates_norm_corr} \small
(a) Least squares correlation coefficient between empirical and MLE-based ideal Fisher--Snedecor CDFs, as a function of frequency.
(b) Normalization constant $a=C/\langle Q \rangle$ for curve fit to CDF for MLE estimated values of $M$ shown in Fig. \ref{fig:Mestimates_MoM_MLE}, as a function of frequency, based on $n$-point frequency averaging at each one of $1400$ data points $q$ for $Q$.}
}
\end{figure}

Fig. \ref{fig:Mestimates_MoM_MLE} shows results for the estimates of the scaled spectral density of $M$, defined as
\bea
{\cal M}(f) \stackrel{\Delta}{=} M(f) / n
\eea
in units $
(150~{\rm kHz})^{-1}$, obtained by the MoM and MLE. Both methods are seen to produce very similar haracteristics.
For $n > 1$, 
$
{\cal M}(f)$
converges towards $1$ as frequency increases, indicating that each frequency asymptotically contributes one DoF in $f_Q(q)$ when using spectral averaged values of $q_s$.
The further $n$ is increased, the more rapidly ${\cal M}(f)$ approaches $1$. Increasing $n$ also further improves the agreement between the ideal Fisher--Snedecor theoretical and the empirical CDFs, which manifests itself by the correlation coefficient $\rho$ between empirical and estimated CDFs approaching $1$ still better (Fig. \ref{fig:Mestimates_norm_corr}(a)) and, indirectly, by a similarly rapid approach to $1$ for the extra normalization constant $a$ of the fitted CDF (Fig. \ref{fig:Mestimates_norm_corr}(b)), which accounts for any possible deviation from the normalization of $Q$ and its PDF.
The improved accuracy with increasing $n$ is intuitive clear from the description in Sec. \ref{sec:method}, because a wider bandwidth implies more participating cavity modes, which lead to closer agreement between spatial and spectral averages of energy and power, under the assumption of ergodicity to hold.

For $n=1$ (i.e., single-frequency S-parameter values used in the estimation of $q_s$), the estimated $M$ reach an asymptotic value different from $n$ (Fig. \ref{fig:Mestimates_MoM_MLE}).
The lack of averaging then causes the estimation to fail to produce a representative value for the spatial averages $\langle U \rangle_V$ and $\langle P_\rmd \rangle_S$. For $f \gg 1$ GHz, $M$ is approximately $2.5$ on average in this case, 
with a reduced fluctuation spread as frequency increases.
{Unlike for $n>1$, there is no effect of frequency correlation influencing the estimation of ${\cal M}$ when $n=1$ (cf. discussion below): at this value of $n$, the characteristic ${\cal M}(f)$ reflects the performance of the mechanical stirrer, which explains why ${\cal M}(f)$ does then not decrease.}
A distinct discrepancy remains when estimation is based on a single parameter $M$, as $\rho$ is close to, but no longer converging toward $1$. 

\subsection{Estimation of $r$ and $s$}
\begin{figure}[!ht] \begin{center} \begin{tabular}{c}
\ \epsfxsize=8.6cm \hspace{-1cm}
\epsfbox{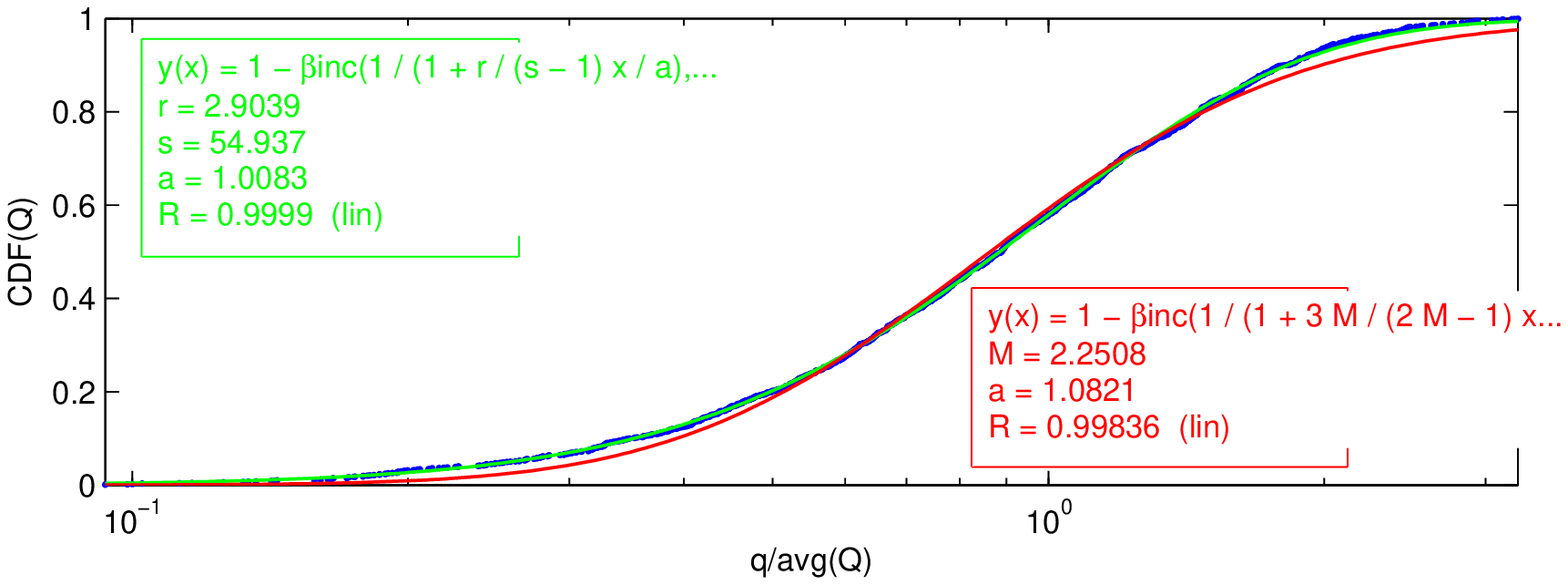}\ \\
(a)\\
\ \epsfxsize=8.6cm \hspace{-1cm}
\epsfbox{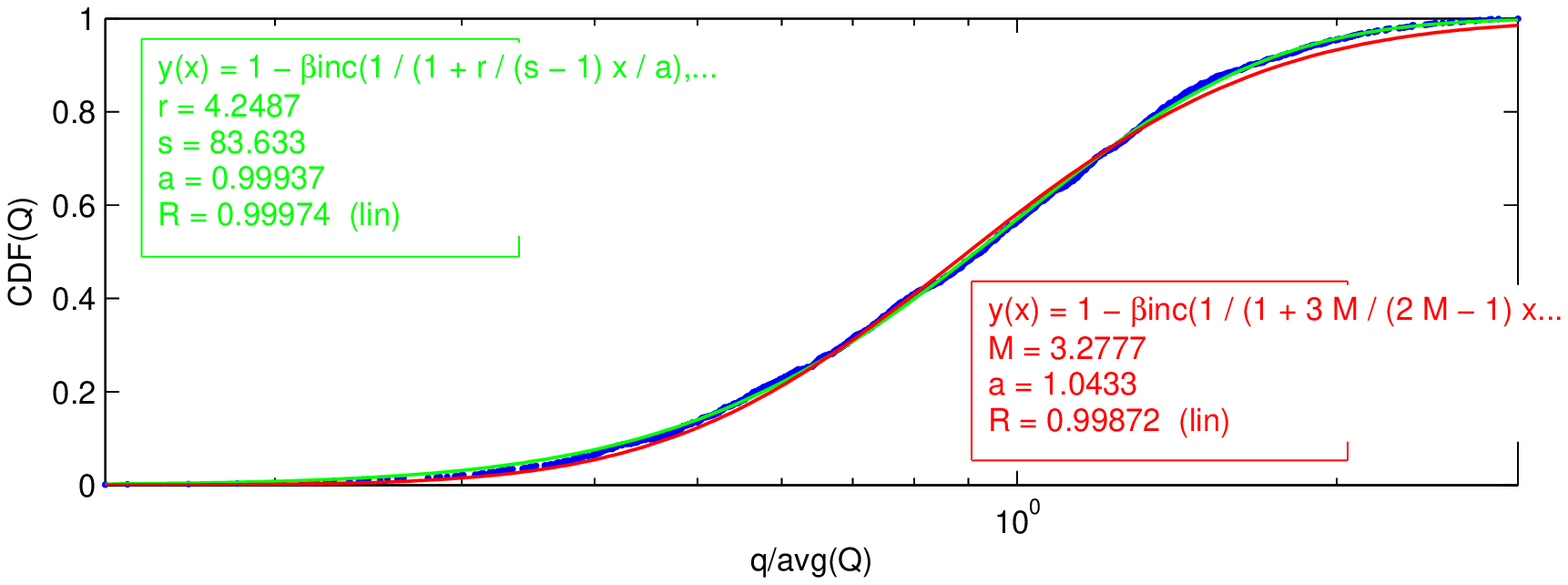}\ \\
(b)\\
\ \epsfxsize=8.6cm \hspace{-1cm}
\epsfbox{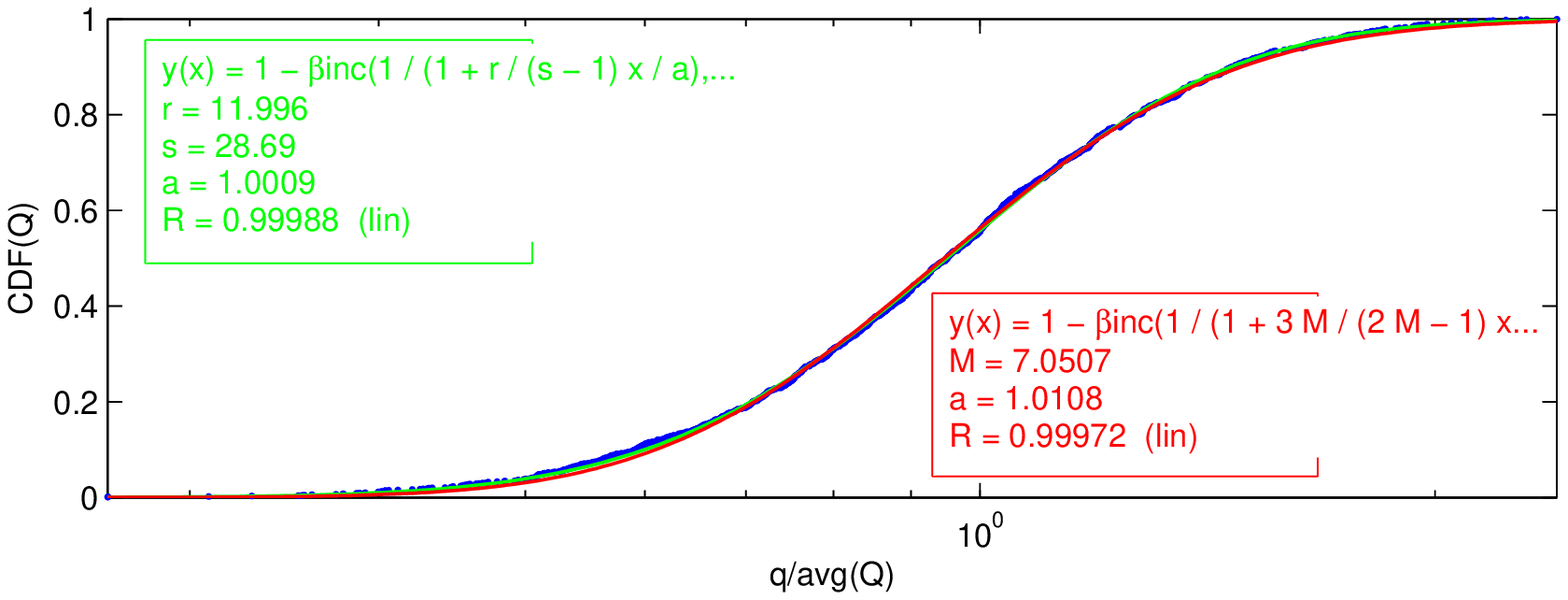}\ \\
(c)\\
\ \epsfxsize=8.6cm \hspace{-1cm}
\epsfbox{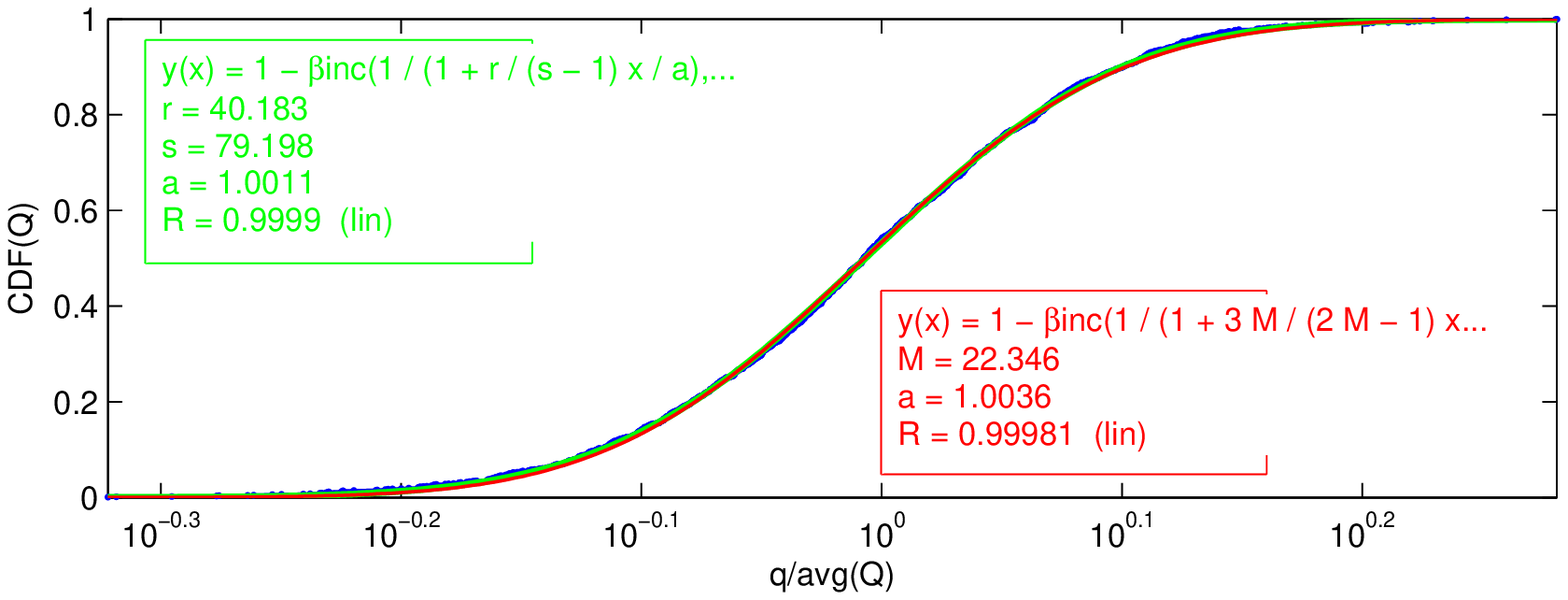}\ \\
(d)
\end{tabular}
\end{center}
{
\caption{\label{fig:theovsemp_curvefit_n1_f5p95GHz} \small
Comparison of empirical CDF (blue dotted) and theoretical Fisher--Snedecor CDFs (red and green lines; eq. (\ref{eq:PDFQ_selfsuff_mu})) for $Q$ at $f=5.95$ GHz, obtained with ($n > 1$) or without ($n = 1$) spectral averaging across 
averaging bandwidth $\Delta f = 150(n-1)$ kHz: (a) $n=1$, (b) $n=2$, (b) $n=5$, and (d) $n=20$ sample frequencies. 
Corresponding values of $r$, $s$, $M$ and correlation between modelled and empirical CDFs quantified by the coefficient of determination $R \stackrel{\Delta}{=} \varrho^2$ (quality of fit) are listed in the insets.
}
}
\end{figure}

When the NDoFs $r$ and $s$ are permitted to be estimated independently by removing the constraints $r=3M$ and $s=2M$ for the distributions of the stored energy and dissipated power, respectively, then a still closer fit can be achieved with the Nelder-Mead algorithm. However, $s \gg 2M$ and $r \ll 3M$ are found.
As an example, typical discrepancies between theoretical and empirical CDFs $F_Q(q)$ for $n=1,2,5,20$ at $f=5.95$ GHz are shown in Fig. \ref{fig:theovsemp_curvefit_n1_f5p95GHz} (note the logarithmic scale for the abscissae). This Figure suggests that the empirical CDF (blue dotted curve) has a more rapid rise and more rapid decay of probability density ($f_Q(q)=dF_Q(q)/dq$), resulting in a slightly more sharply localized density compared to the one-parameter fitted Fisher--Snedecor model (red line curve).
The individually estimated NDoF $r$ for the stored energy has a comparatively small value (e.g., $r = 2.9039 \ll 3 \times 2.2508$ for $n=1$), whereas the corresponding NDoF $s$ for dissipation is significantly higher than for ideal random fields (e.g., $s = 54.937 \gg 2 \times 2.2508$ for $n=1$). 
Thus, despite the excellent agreement between the fitted CDF for independent $r$ and $s$ (green line) and the empirical CDF (blue dotted curve), {\it a fortiori} when $n >1$, this discrepancy suggests that these values of $r$ and $s$ are not representative NDoFs for the actual fields and that spectral averaging is essential in order to achieve a reasonable degree of accuracy of the model.

\subsection{Effect of Modal Coupling}
Returning to Fig. \ref{fig:Mestimates_norm_corr}, confirmation of the increasing accuracy of the theoretical Fisher--Snedecor PDF (\ref{eq:PDFQ_selfsuff_mu}) with $r=3M$, $s=2M$ is demonstrated by the fact that the normalization constant $C/\langle Q \rangle$ and correlation coefficient $\rho$ for the matching of CDFs approach unity when $n$ and $f$ increase.
For $n>1$, the reason for the initial increase of $M/n$ at low and intermediate frequencies, followed by a decrease toward $1$ is not obvious and can be understood as follows. At higher frequencies, the probability of having two (or more) modes with smaller spectral separation $\delta_{mnp}$ is obviously higher than at lower frequencies, on account of the Weyl law for the modal density. As a result, intermodal coupling is more likely to occur at higher frequencies because of stronger modal overlap (nonzero resonance bandwidths), which positively affects the randomization of the field. Because of this overlap, some measure of modal correlation -- in the sense of the excitation of one mode causing the partial excitation of another adjacent one -- then causes a saturation of the number of {\em independent\/} modes and, hence, of the NDoF governing the field.

\begin{figure}[!ht] \begin{center} \begin{tabular}{c}
\ \epsfxsize=6cm 
\epsfbox{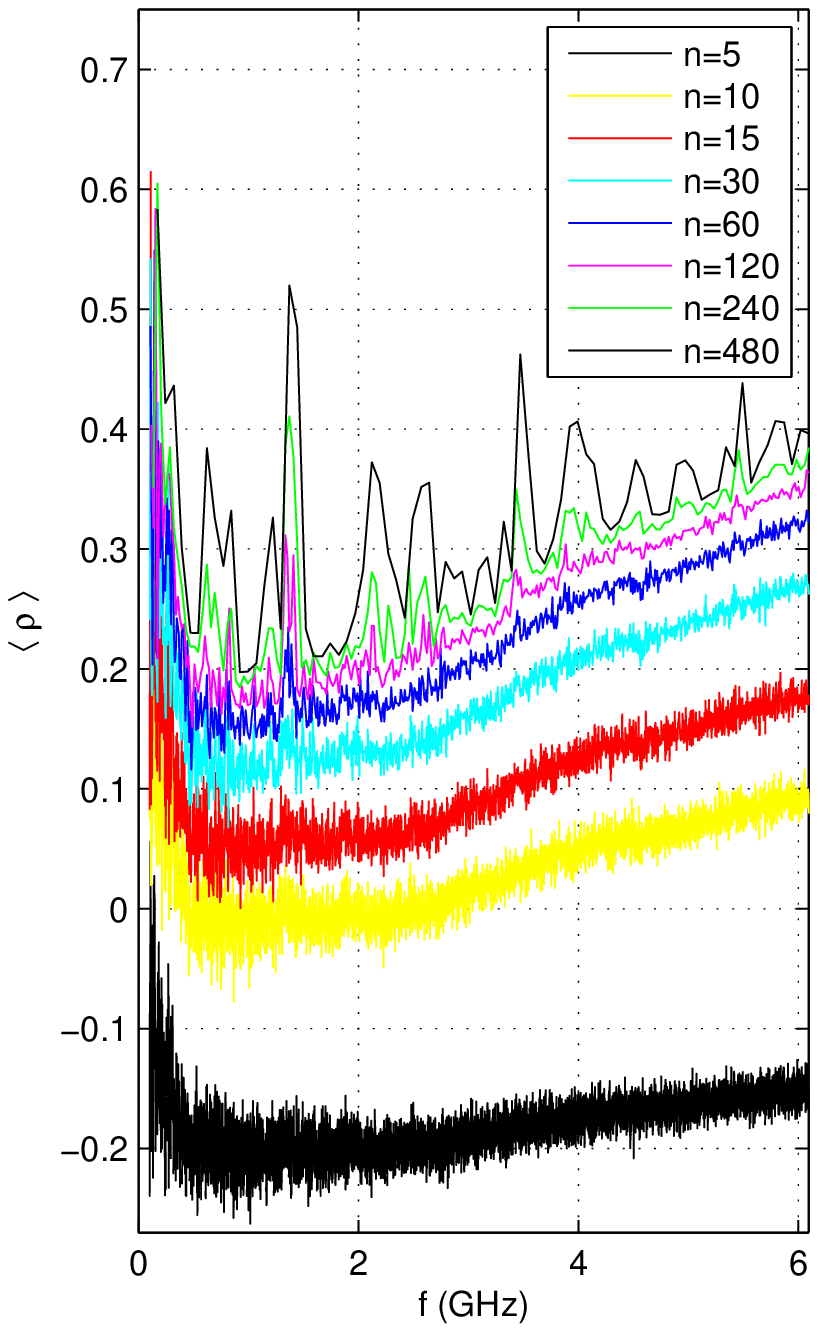}\
\\(a) \\
\ \epsfxsize=6cm 
\epsfbox{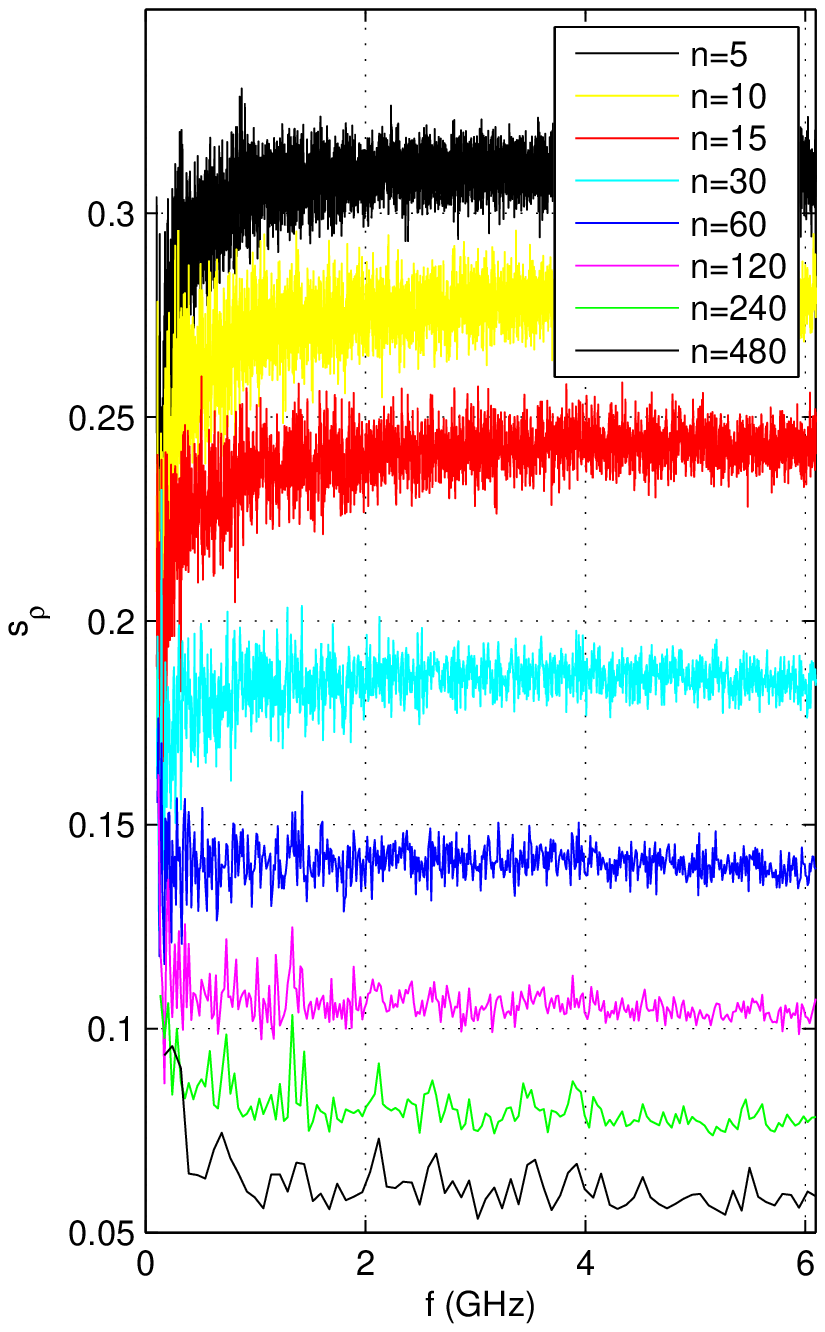}\ \\
(b) \\
\end{tabular}
\end{center}
{
\caption{\label{fig:ACF_rho} \small
(a) Sample average $\overline{\rho}$ 
and
(b) standard deviation $s_\rho$ of one-step frequency correlation coefficient $\rho$ for $|S_{21}(f)|^2$ measured at $1400$ stir states, calculated for frequency subintervals $\Delta f = (n-1)\delta f$ with $\delta f = 150$ kHz.
}
}
\end{figure}

{To investigate this effect more closely, we calculated the sampled frequency\footnote{To show the effects of frequency dispersion, this was done for different finite subintervals (sample size, record length) and for frequency blocks of different size $n \Delta f$, with $n$ ranging from $5$ to $480$. Note that $n$ cannot be chosen too large, even though this reduces $s_\rho$, if one wishes to investigate the frequency dependence of $\overline{\rho}$. For $n=2$ and $3$, $\overline{\rho} = -1$ and $-0.5$, respectively.} correlation coefficient $\rho(\tau)$ between the measured frequency spectrum $|S_{21,\alpha}(f)|^2$ (Cartesian field, single antenna orientation) and its replica shifted by one frequency step, i.e., $|S_{21,\alpha}(f+150~{\rm kHz})|^2$,
 for each stir state $\tau$, followed by averaging of the obtained sampled coefficients over all $1400$ stir states to yield their sample average $\overline{\rho(f)}$ and associated standard deviation $s_\rho(f)$. 
Fig. \ref{fig:ACF_rho}(a) shows $\overline{\rho(f)}$ for selected values of $n$. The data confirm an initial sharp decrease in $\overline{\rho(f)}$ in undermoded regime -- corresponding to a sharp increase in the number of statistically independent frequencies or cavity modes $N\propto 1/\overline{\rho}$ as mode count increases without significant modal overlap -- which is then followed by $\overline{\rho(f)}$ increasing approximately linearly above $1$ GHz as modal overlap takes effect. The accompanying decrease of $s_\rho(f)$ for large $n$ shown in Fig. \ref{fig:ACF_rho}(b) indicates that this frequency dependence of $\overline{\rho(f)}$ becomes statistically increasingly significant.
}

The behaviour of ${\cal M}(f)$ is not only important for the modelling of the PDF of $Q$, but it also relates directly to the mean value $\langle Q \rangle$ and its frequency dependence, because of the result \cite[Eqs. (18) and (21)]{arnaTEMC2013}
\bea
\langle Q(f) \rangle = \left ( 1 + \frac{1}{2M(f)} \right ) Q_\infty(f).
\eea
Since $Q_\infty(f) \propto \sqrt{f}$, a physical reason for the saturation of the value of $M(f)$ for a selected value of $n$ would physically explain the fact that the experimentally observed value of $\langle Q(f) \rangle$ becomes saturated in highly overmoded conditions.
Evidently, additional reasons (e.g., the effect of surface roughness and other nonideal operational or measurement conditions) may contribute to such saturation.

\subsection{Number of Degrees of Freedom per Frequency}
\begin{figure}[!ht] \begin{center} \begin{tabular}{c}
\ \epsfxsize=8.6cm \hspace{-0.6cm}
\epsfbox{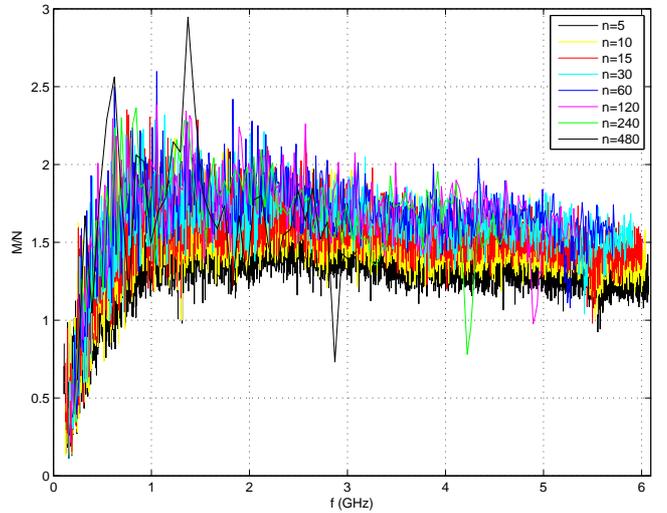}\
\end{tabular}
\end{center}
{
\caption{\label{fig:MoverNestimation} \small
Ratio $M(f)/N(f)$, 
based on MLE estimated $M(f)/n$ (cf. Fig. \ref{fig:Mestimates_MoM_MLE}(b)) and $\overline{\rho(f)}$ (cf. Fig. \ref{fig:ACF_rho}(a)).
}
}
\end{figure}

{From the frequency characteristics of $M/n$ and $\rho$, the ratio of the NDoF, $M$, to the number of `equivalent' independent frequencies, $N$, can be approximately estimated as follows. For $\rho=0$, all $n$ sampled frequencies are independent whence $N=n$. At the other extreme, for $\rho=1$ all measurement frequencies within a block are perfectly correlated, leading to $N=1$. 
Assuming a simple linear\footnote{Exponential or other more sophisticated models can also be adopted.} model for the dependence of $N$ on $\rho$ , this yields $N=n+(1-n)\rho$. Thus, 
\bea
\frac{M}{N} \simeq \frac{M}{n+(1-n)\rho}.
\label{eq:MoverN}
\eea
Fig. \ref{fig:MoverNestimation} shows this ratio as a function of frequency, based on this correlation model with the data of Figs. \ref{fig:Mestimates_MoM_MLE}(b) and \ref{fig:ACF_rho}(a). 
}

\subsection{Empirical vs. Theoretical Distribution of $Q$}
\begin{figure}[!ht] \begin{center} \begin{tabular}{c}
\ \epsfxsize=8.6cm \hspace{-0.6cm}
\epsfbox{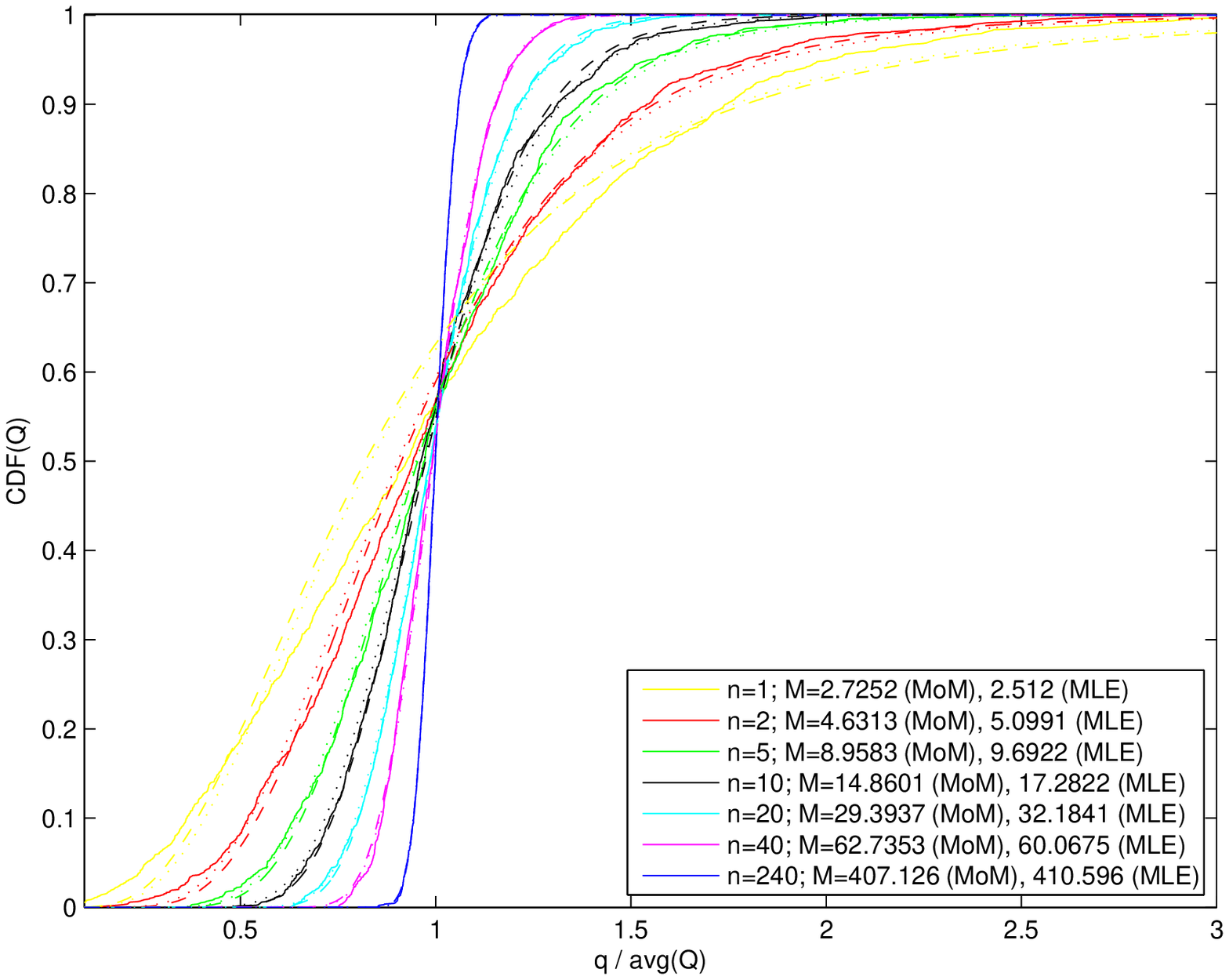}\ \\
(a)\\
\ \epsfxsize=8.6cm \hspace{-0.6cm}
\epsfbox{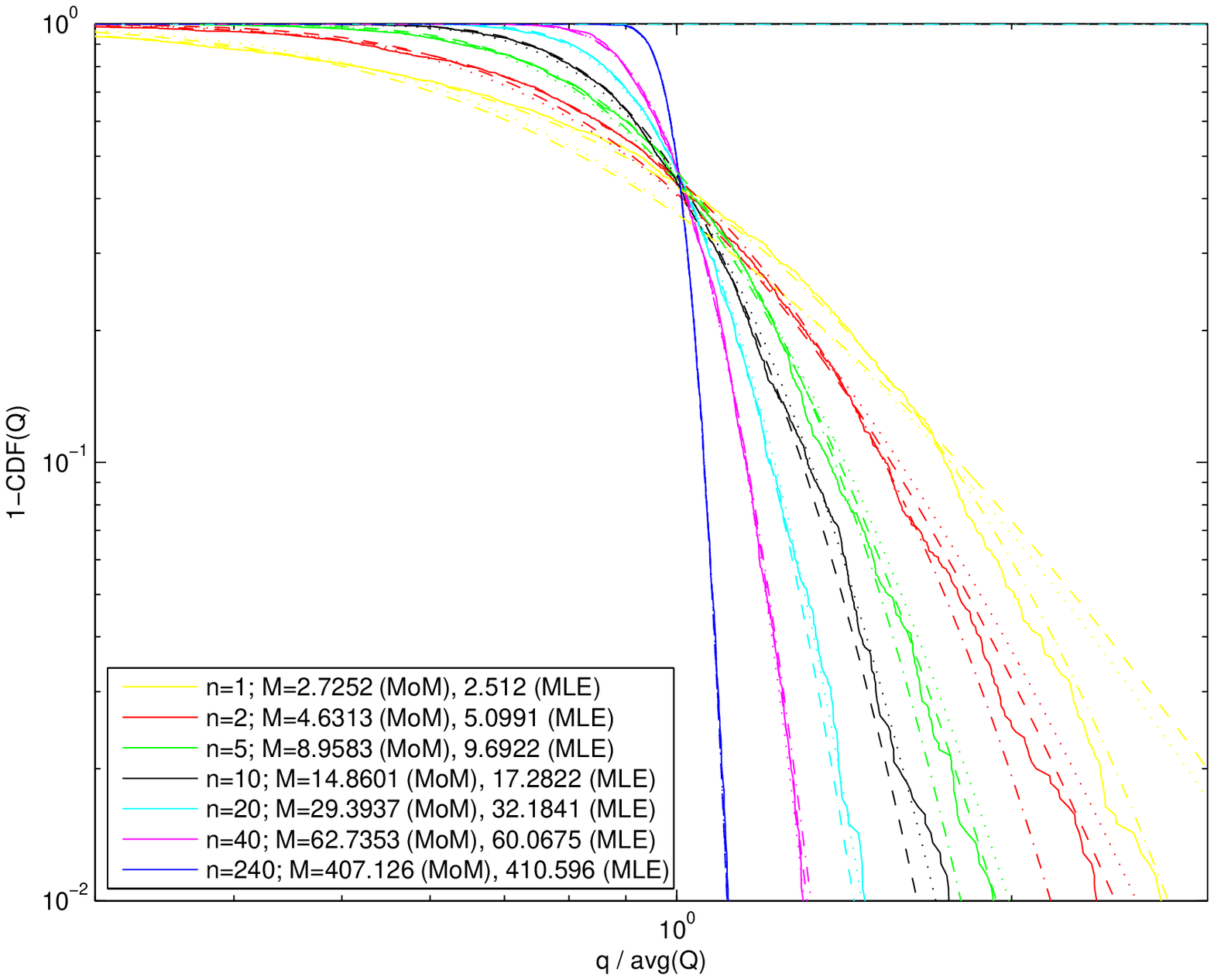}\ \\
(b)\\
\end{tabular}
\end{center}
{
\caption{\label{fig:CDF_F1GHz_param_n} \small
(a) CDF and (b) complementary CDF of $Q$ using spectral averaging method applied across $n$ sample frequencies (bandwidth $\Delta f = 150(n-1)$ kHz) centered at $1$ GHz, for selected values of $n$. 
Dotted: theoretical Fisher--Snedecor CDF (\ref{eq:PDFQ_selfsuff_mu}) based on empirical MoM estimated parameter $M$.
Long-dashed: theoretical Fisher--Snedecor CDF  with empirical MLE parameter $M$.
Solid: empirical CDF (raw data). 
}
}
\end{figure}

Fig. \ref{fig:CDF_F1GHz_param_n} compares the measured and theoretical CDFs and complementary CDFs at an arbitrarily chosen frequency ($1$ GHz), for selected values of $n$ and based on the estimated values for $M$ that were shown in Fig. \ref{fig:Mestimates_MoM_MLE}.
It can be seen that even for values of $n$ as low as $5$, a fairly good agreement is already achieved that rapidly improves further when the frequency $f$ or the spectral averaging $n$ is increased.

\section{Conclusion\label{sec:conclusions}}
In many studies, the $Q$ of a MT/MSRC often appears as a deterministic quantity with a fixed
constant (average, effective) value for simplicity. The present paper has demonstrated that the spread on values of $Q$ across stir states can be considerable when the number of simultaneously excited cavity modes is not exceedingly large ($M \not \rightarrow +\infty$), as is typical in normal (narrowband) operation of a mechanically or electronically stirred chamber. Therefore, the full PDF of $Q$ needs to be taken into account for accurate theoretical uncertainty quantification, as well as in calculation of the measurement uncertainty budget in experiments. 

In this paper, we proposed and applied a spectral averaging method for estimating the PDF of $Q$. The technique requires measurements of S-parameters only at a single pair of antenna locations, across a set of stir states and across a narrow frequency interval to enable spectral averaging. The data for these local measurements provide accurate estimates of the ratio of the stored energy across the entire cavity volume to the ohmic dissipated power dissipated in the walls, per stir state. Repeating for a full set of stir states, this yields corresponding estimates of the distribution of $Q$.

The spectral averaging methods avoids the need for elaborate spatial scanning of the antennas to estimate the stored and dissipated energy at each stir state. Also, the practical difficulties and inaccuracies of measuring the dissipated power based on field measurements near a conducting boundary are thus circumvented.

The measurement results confirm the viability and accuracy of the spectral averaging method for determining $f_Q(q)$ when used with mechanical stirring.
The technique is highly efficient in the GHz range of a typical chamber: typically, only about $n=20$ spectrally independent values ($\sim \Delta f = 3$ MHz averaging bandwidth) are required to get excellent agreement, while even smaller averaging bandwidths (e.g., $n=3 \sim \Delta f = 300$ kHz) still produce accurate statistics of $Q$ at $1$ GHz, except for its extreme values.   
It was also shown that a simple moment method estimation of $M$ is very close to a more elaborate maximum likelihood estimation in all cases investigated.

The scaled spectral density 
$M(f)/[150(n-1)]$ tends to some universal behaviour. It rapidly increases with frequency below $1$ GHz, followed by a slow decrease towards $1$ above $1$ GHz. The latter is explained by a saturation of modal population, which occurs when the mode density and modal overlap increase causing increased statistical dependence among closely spaced excited modes.

\section{Acknowledgement}
L.R.A. wishes to thank P.B. and INSA Rennes for their invitation and hospitality during the period in which this work was performed.

\end{document}